\documentclass[aps,twocolumn]{revtex4-1}
\usepackage{epsfig}
\usepackage{color}
\usepackage{amsmath}
\usepackage[retainorgcmds]{IEEEtrantools}

\usepackage{lipsum}

\begin{document}

\title{Influence of distinct kinds of temporal disorder
  in discontinuous phase  transitions }

\author{Jesus M. Encinas and 
C. E. Fiore$^1$}
\email{fiore@if.usp.br} 

\affiliation{$^1$ Instituto de F\'{\i}sica,
Universidade de S\~{a}o Paulo, \\
Caixa Postal 66318\\
05315-970 S\~{a}o Paulo, S\~{a}o Paulo, Brazil }
\date{\today}

\begin{abstract}
Based on the MFT arguments, a general description for
discontinuous phase transitions in the presence
temporal disorder is considered.
   Our analysis extends the recent findings [Phys. Rev. E {\bf 98}, 032129 (2018)] by considering
   other kinds of phase transitions beyond the absorbing ones.
 The theory is exemplified in the simplest   (nonequilibrium) order-disorder (discontinuous) phase transition with "up-down" $Z_2$ symmetry: the inertial majority vote (IMV) model
 for two kinds of temporal disorder.
As
for the APT case, the temporal disorder does not suppress
the occurrence of discontinuous phase transitions, but remarkable differences emerge when compared with the pure case. A comparison between the distinct  kinds of temporal disorder is also
performed beyond the MFT for random-regular (RR) complex topologies.
\end{abstract}

\maketitle

\section{Introduction}

Disorder due  inhomogeneities is  present
 in many real systems
and commonly plays a significant role in their behaviors \cite{marr99,henkel,odor07}. In the
last years, many attention has been devoted
for critical phase transitions in the presence of spatial \cite{vojta05, vojta06,  oliveira08,vojta09,vojta12} and temporal disorders \cite{munoz2011,martinez,vojta-hoyos,vojta16b,neto2,solano,temp}, in which one has established the existence of  new (and universal) critical behaviors. Remarkably, both kinds of disorder are also characterized by the existence
 a subregion of phase space in which one observes  exotic behaviors. The former is named spatial 
Griffiths phase and corresponds to  a subregion in the absorbing phase in which
the   order parameter vanishes slower than (power-law or  stretched exponential) the exponential
decay in the absence of disorder. Conversely,
 temporal disorder is featured by a region in the active phase in which the mean lifetime increases as a power-law  (instead of exponential).

Spontaneous breaking symmetry manifests in a countless 
sort of systems beyond the classical
ferromagnetic-paramagnetic phase transition \cite{marr99,henkel}.
It includes remarkable examples,
 such as
school fishes moving under an ordered way for protecting
themselves against predators, 
spontaneous formations of a common language and culture, 
the emergence of consensus \cite{vicsek,aceb,loreto} in social systems and other remarkable examples.
Such phase transitions   are typically critical and belong to  well established universality classes \cite{marr99,henkel,mario92,odor07}. A remarkable 
example  commonly considered  for modeling/describing some of above phenomena is the majority vote (MV) model \cite{mario92,chen1,pereira},
in which a local spin tends  to align itself with its 
 local neighborhood majority
spins. Originally, it presents a continuous
phase transition belonging to distinct universality classes, according to the lattice topology \cite{mario92,chen1,pereira}.
More recently \cite{chen2,harunari,jesus},
it has been found
that the inclusion of inertia in
the MV (IMV), e.g.  a term proportional to the
local spin, can shift the phase transition  to
 a discontinuous
 phase transition in complex networks \cite{chen2,harunari}  or even
 in regular lattices \cite{jesus,fsize2}. 
 The importance of such results is highlighted by the fact that
 behavioral inertia is an essential
 characteristic of human being and animal groups and it is also
 a significant ingredient triggering abrupt transitions
that arise in social systems \cite{loreto}. 
However, the effects under
the inclusion of more realistic ingredients,
such its  time dependent variation,
have not been  satisfactorily understood yet.

Recently, a theory for discontinuous absorbing  phase transitions (APTs) in the presence of temporal disorder has been proposed in Ref. \cite{fioreneto}. In contrast to the spatial disorder case \cite{paula}, discontinuous APTs are not suppressed
due to the temporal disorder, although remarkable
features emerge when compared with their pure (disorderless) systems.
Giving that  systems with $Z_2$ "up-down" symmetry  display remarkably different features from APTs (which can be viewed in
terms of the simple
logistic  order-parameter $x$ equations: $dx/dt=ax-bx^2+cx^3...$ (APT) and $dx/dt=ax-bx^3+cx^5...$ ($Z_2$) \cite{app}), a  question that naturally arises is if
similar findings are verified/can be extended
beyond the APT. 
Additionaly, a second important point concerns at a comparison between distinct kinds of
temporal disorder. In the specific case of IMV, giving that the inertia plays
a fundamental role for shifting the phase transition, we intend
to tackle the effect of temporal
disorder in the inertia and their differences respect to 
 the (usual) control parameter case.

Aimed at answering
aforementioned points, here we examine,
separately, the role of temporal disorder in two
fundamental ingredients: the control parameter and inertia. Based on mean-field analysis, we derive general predictions for both kinds of
temporal disorder, which are also
verified beyond in the MFT for complex structures.

This paper is organized as follows: In Sec. II we present the analysis of pure model and temporal disorder based on the MFT, in Sec. III we present the main findings beyond the MFT. Conclusions are drawn in Sec. IV.

\section{Model and mean field analysis}
The original (inertialess)
majority vote model (MV) is defined as follows.  At each  time step, a site $i$ with spin $\sigma_i$ is
randomly selected and 
with probability $1-f$ it is aligned with the majority of its $k_i$
nearest-neighbors and with the complementary probability $f$
the majority rule is not followed. The inertial majority vote model (IMV)
differs from the MV for the inclusion of an inertial term $\theta$,
taking into account the contribution of the local spin. In such a case,
the probability of following the majority rule will also depend
on the local spin $\sigma_i$, whose transition rate $\omega_i(\sigma)$ from $\sigma_i \to -\sigma_i $ is given by \cite{chen2} 
\begin{equation}\label{eq1}
\omega(\sigma_i) = \frac{1}{2}\left[1-(1-2f)\sigma_iS(\Theta_i)\right],
\end{equation} 
where $\Theta_i$ accounts for the local neighborhood plus the inertial contribution given by  $$\Theta_i = (1-\theta)\sum_{j=1}^{k_i} \frac{\sigma_j}{k_i} + \theta\sigma_i,$$
with $S(x)={\rm sign}(x)$ if $x\neq0$ and $S(0)=0$. Note that one recovers
the original MV when $\theta=0$ and an order-disorder
phase transition yields only when
the inertia  is constrained between
$\theta \in [0,0.5]$. For $\theta=0.5$ the system gets frozen in the order/disorder phase according to whether $f=0/f\neq 0$. 
By increasing $\theta$ and the  connectivity, phase transition
is shifted  from a continuous (second-order)  to a  discontinuous (first-order). At the mean-field
level the phase coexistence is marked by the appearance  of an hysteretic
region in which two symmetric ordered and a 
 disordered phases coexist. Such features are also manifested in complex networks but an entirely different behavior is presented for regular lattices \cite{jesus}, in which quantities scale with the system volume.

From the transition rate, the time evolution of  the average magnetization
$m_k=\langle\sigma_i\rangle_k $ of a
local site $i$ with degree $k$ is given by  
\begin{equation}\label{eq2}
\frac{d}{dt}m_k = - m_k + (1-2f)\langle S(\Theta_i)\rangle.
\end{equation}
The first analysis will be performed by means of a
MFT treatment, in which the joint probabilities 
appearing in 
the average  $\langle S(\Theta_i)\rangle$
are a rewritten
in terms of one-site probabilities. From this assumption, one arrives the following expression $\langle S(\Theta_i)\rangle=(1+m_k)\langle S(\Theta_+)\rangle/2+(1-m_k)\langle S(\Theta_-)\rangle/2$,
where $\langle S(\Theta_\pm)\rangle$ are
given by
\begin{equation}\label{eq3}
\langle S(\Theta_\pm)\rangle \approx \sum_{n=\lceil n_k^{\pm} \rceil}^k C_n^k p_+^np_-^{k-n}-\sum_{n=\lceil n_k^{\mp} \rceil}^k C_n^k p_-^np_+^{k-n},
\end{equation}
with $p_{\pm}$ being the probability that a nearest neighbor is
$\pm 1$ (in which one associates 
the local magnetization $p_\pm=(1\pm m^*)/2$) and $n_{k}^{-}$ and $n_{k}^{+}$  correspond to the lower limit 
of the ceiling function given by $n_{k}^{-}=k/[2(1-\theta)]$
and $n_{k}^{+}=k(1-2\theta)/[2(1-\theta)]$, respectively..

In order to relate
$m^*$ and $m_k$, we shall focus our analysis on  uncorrelated networks, in
which the probability 
of a randomly chosen
site has degree $k$ reads $kP(k)/\langle k \rangle$, with $P(k)$ and $\langle k \rangle$
being the probability distribution of nodes and its
mean degree $\langle k \rangle$, respectively.  The relation between
$m^*$ and $m_k$ then reads
$m^* = \sum_k m_k kP(k)/\langle k \rangle$.
By combining above expression with Eq. \eqref{eq2}, we  obtain the
following  self-consistent
equation of $m^*$ in the steady-state regime:
\begin{equation}\label{eq5}
\scriptsize
m^*=(1-2f)\sum_{k}\frac{kP(k)}{\langle k \rangle}\left[\bigg(\frac{1+m_k}{2}\bigg) \langle S(\Theta_+)\rangle + \bigg(\frac{1-m_k}{2}\bigg) \langle S(\Theta_-)\rangle \right]
\end{equation}
Above expression can be analyzed for distinct complex structures. For a 
 random-regular (RR) topology, $P(k)$ is given by $P(k)=\delta(k-k_0)$ 
and hence all sites have the same number of
neighbors $k_0$, from which one  arrives at the following expression for
the steady $m^* = m(k_0) \equiv m$
in terms of $f$ and $\theta$:
\begin{equation}\label{eq6}
\footnotesize
m=(1-2f)\left[\bigg(\frac{1+m}{2}\bigg) \langle S(\Theta_+)\rangle + \bigg(\frac{1-m}{2}\bigg) \langle S(\Theta_-)\rangle \right].
\end{equation}

Above expressions present two and three stable solutions in the case of continuous and
discontinuous phase transitions, marked at $f=f_c$ (critical point) and $f=f_f$ (order-parameter jump), respectively. In both cases,  
there is a trivial  solution, $f>f_c$ $(f>f_f)$ corresponding to the disordered (DIS) phase: $m(t\to\infty) = m_d(f)=0$, 
 irrespectively on the initial condition. 
Conversely, for $f<f_b$, 
the system evolves to  $m(t\to \infty) \to m_s(f)$, also independently on the initial
condition.  
The third solution $m_u(f)$ is called unstable solution and appears for values of $f$ constrained in the 
interval $f_b<f<f_f$. More specifically, $m(t\to\infty) \to m_s(f)$ and $m(t \to \infty) \to m_d(f)$,
if $m(0)>m_u(f)$ and  $m(0)<m_u(f)$, respectively. This feature of the ordered phase  will be refereed as the metastable (ME) phase, contrasting
with the behavior for $f<f_b$. Although
the discontinuous phase transition yields at $f=f_f$, the region $f=f_b$ marks  the crossover
between an ordered phase characterized by bistable  and monostable behaviors for $f_b<f<f_f$ and $f<f_b$, respectively.
Since $m(t)$ deviates from $m_u(f)$ whenever $m(0)\neq m_u(f)$, such a solution is unstable.

 Although  analytic expressions and the stability of solutions based on Eqs. (\ref{eq5}) and (\ref{eq6}) 
  are quite cumbersome, a simpler analysis can be 
performed in the limit of large connectivities,  since   each term of the binomial distribution
  approaches  a Gaussian
  with mean $kp_{\pm }$ and  variance $\sigma^{2}=kp_{+}p_{-}$ \cite{chen1,romualdo,harunari,noa,jesus2}. From Eq. (\ref{eq6}),  the first term from the right side is approximately rewritten as 
 \begin{eqnarray}&&\sum_{n=\lceil n_k^\pm \rceil}^{k_0}C_{n}^{k_0}p_{\pm}^{n}p_{\mp}^{k_0-n} \rightarrow  \frac{1}{\sigma\sqrt{2\pi}}\int_{n_k^+}^{k_0} e^{-\frac{(\ell-k_0p_{\pm})^2}{2\sigma^2}}d\ell=\nonumber\\&=&
\frac{1}{2}\sqrt{\pi}\left\{{\rm erf}\left[\frac{k_0(1-p_{\pm})}{\sqrt{2}\sigma}\right]-{\rm erf}\left[\frac{k_0(n_k^+-p_{\pm})}{\sqrt{2}\sigma}\right]\right\},
\end{eqnarray} 
with ${\rm erf(x)}$ denoting the error function ${\rm erf(x)}=2\int_{0}^{x}e^{-t^2}dt/\sqrt{\pi}$,
and  the second one can be rewritten under a similar way. 
Taking into account that ${\rm erf}[k_0(1-p_{\pm})/\sqrt{2}\sigma]$ approaches
to $1$ for large $k_0$, we arrive at the following
expression for the steady state regime:
\begin{equation}
f=\frac{1}{2}\left[1-\frac{2m}{(1+m){\rm erf(\alpha)}-(1-m){\rm erf(\beta)}}\right],
\label{eq62}
\end{equation}
with parameters $\alpha$ and $\beta$  given by 
\begin{equation}
\alpha=\sqrt {\frac{k_0}{2}}\left[\frac{\theta}{1-\theta}+m\right] \quad {\rm and} \quad \beta=\sqrt {\frac{k_0}{2}}\left[\frac{\theta}{1-\theta}-m\right].
\end{equation}
The transition point $f_f$
can be obtained from the maximum of Eq. (\ref{eq62}).
At the vicinity of  
$f_b$ (or $f_c$ for a critical phase transition), $m$ is expected to be small and hence one has
the following  logistic equation  $dm/dt \approx A(f,\theta,k_0)m$, with $A(f,\theta,k_0)$  given by
\begin{equation}
\footnotesize
A(f,\theta,k_0)=-1+(1-2f)\left[\sqrt{\frac{2k_0}{\pi}}e^{-\frac{k_0\theta^2}{2(1-\theta)^2}}+{\rm erf}\left(\sqrt\frac{k_0}{2}\frac{\theta}{1-\theta}\right)\right].
\label{largek}
\end{equation}
From the above expression, $f_b$ is 
then given by $A(f_b,\theta,k_0)=0$. Note
that one reduces to the expression $2f_c=1-\sqrt {\pi/(2k)}$, when $\theta=0$ \cite{chen1,noa}.
Hence, $m(t)$ is exponentially increasing
towards
its steady state value $m_s(f)$ if $A(f,k_0,\theta)>0$ $(f<f_b)$ 
and vanishes exponentially for $A(f,k_0,\theta)<0$, $(f_b<f<f_f)$, respectively, if $m(0)<<1$   [see e.g. Figs .\ref{fig1}$(c)$ and $(d)$(inset)].
 Since $f_f>f_b$, 
 $m(t)$
also vanishes exponentially towards
$m_d(f)$.

In order to illustrate all previous findings, Fig. \ref{fig1} depicts, for the clean system, the phase diagram and all above main features 
of dicontinuous phase transitions for $k_0=12$ and $\theta=0.45$ as $f$ is changed.
In particular, the regions $f\le f_b=0.0274573...$ and $f_b<f<f_f=0.080121$ mark  the ORD and ME phases, respectively, whereas 
for $f> f_f$ the disordered phase (DIS) prevails.
Similar results are obtained for other 
connectivities $k_0$ and $\theta$.
\begin{widetext}
\begin{center}
\begin{figure}
\includegraphics[scale=0.6]{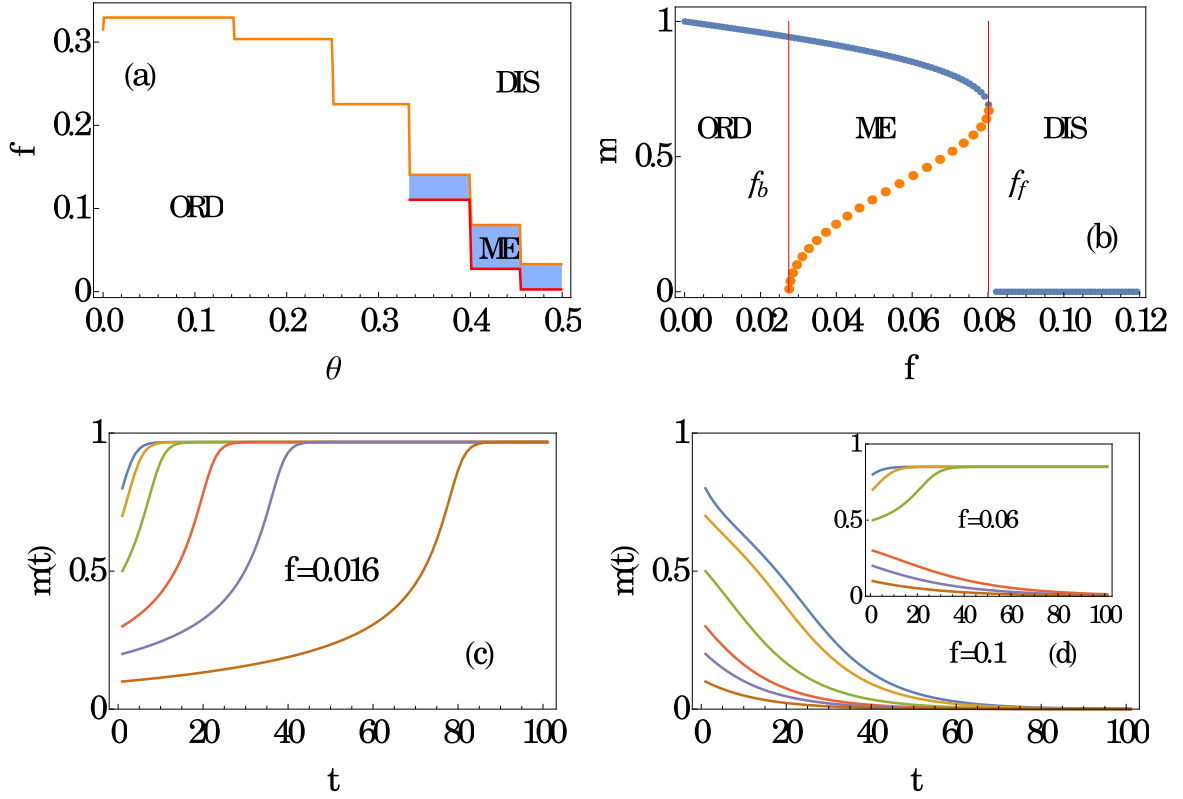}
\caption{Panel $(a)$ depicts the MFT phase diagram for a RR
topology with $k_0=12$. ORD/ME and DIS denote the ordered monostable/bistable and disordered phases, respectively. In $(b)$, the behavior of average magnetization $m$
  versus $f$ for $\theta = 0.45$. Continuous and dashed lines denote to the stable
  and unstable solutions of Eq. (\ref{eq6}),
  respectively.  Their main features are exemplified in panels
  $(c)-(d)$ by taking the time evolution of $m$ as a
  function of time for $f=0.016 (c)$, $f=0.10(d)$ and $0.06$ (inset) for distinct initial conditions $m(0)$.}\label{fig1}
\end{figure}
\end{center}
\end{widetext}

   As a final remark, it is worth mentioning that 
although the dependence between
$m$ and $\theta$ is more cumbersome
than with $f$, all previous findings are hold valid
when the inertia is taken
as the control parameter (for fixed $f$).

\subsection{Temporal disorder in the control parameter}
Once presented the main features about the pure system, we now are in position
for tackling the effects of the temporal disorder. We start with time variations of 
the control parameter $f$.  Although  similar findings are expected for distinct temporal
disorder distributions, we
shall consider a simplest
case in which for a given
time interval constrained between $t$ and $t+\Delta t$, 
control parameter $f$ is randomly extracted from a bimodal distribution $P_{dis}(f)$:
\begin{equation}
P_{dis}(f)=p\delta(f-f_-)+(1-p)\delta(f-f_+),
\end{equation}
where $f_-<f_+$ and $p(1-p)$ is the probability in which $f$ assumes the values $f_-(f_+)$. During
this time interval,  the system behaves as the pure system, since its control parameter is kept
fixed. For simplicity and also for comparing with previous findings \cite{fioreneto}, 
we set $p=1/2$. 

Analysis starts from a given initial condition $m(0)$ and its time evolution
is analyzed until a sufficient large time $t_{max}$ in which one has generated 
a given sequence of control parameter values $\{f_1,f_2,...,f_M\}$, where $t_{max}=M\times \Delta t$. This process is then repeated for
 sufficiently  $N_D$ distinct disorder sequences (we have considered here $N_D=10^2-10^3$).

Although our findings are not dependent on the value of $k_0$ and $\theta$, the effect of temporal disorder  will be exemplified for $k_0=12$ and $\theta=0.45$, in order to compare
both clean and disordered systems. All possible
variations of both $f_-$ and $f_+$ along ORD, ME and DIS will be considered. 
We face  two scenarios, in which  both $f_-$ and $f_+$
belong to the same and different phases, respectively. 

Let us start with the case when both $f_-$ and $f_+$ varies over the ordered phase $(0\leq f_\pm < f_b)$.  Irrespective on the initial condition $m(0)$ the system will evolve towards an ordered state in which
the steady magnetization fluctuates between $m_s(f_-)$ and $m_s(f_+)$. A similar conclusion is valid for both $f_-$ and $f_+$ belong to the disordered phase $(f_f < f_\pm \leq 1/2)$, in which the disordered phase prevails
independently on $m(0)$. 
For both $f_-$ and $f_+$ belonging to the  metastable phase $(f_b \leq f_\pm \leq f_f$ and $m_u(f_-) < m_u(f_+)$),  then $m(t\to\infty)\rightarrow 0$ and  $m(t\to\infty) \neq0$
if $m(0) < m_u(f_-)$ and  $m(0)>m_u(f_+)$,
respectively,
irrespective the sequence of $f_-$ and $f_+$, respectively. The case in which $m_u(f_-)\leq m(0) \leq m_u(f_+)$  will depend on the particular sequence of $f_-$ and $f_+$.  This can be verified under two extreme cases.
Take for instance a particular (long) sequence of $f=f_+$, in which $m(t)$
becomes lower than $m_u(f_-)$. In such a case,  the system always reaches the disordered phase. Conversely  a long sequence of  $f=f_-$ will lead to  $m(t)>m_u(f_+)$ and then the system will converge to the ordered phase.  Thus, as for absorbing phase transitions \cite{fioreneto},  ORD, ME and DIS phases are preserved under the temporal disorder.
\begin{widetext}
\begin{center}
\begin{figure}
\includegraphics[scale=0.5]{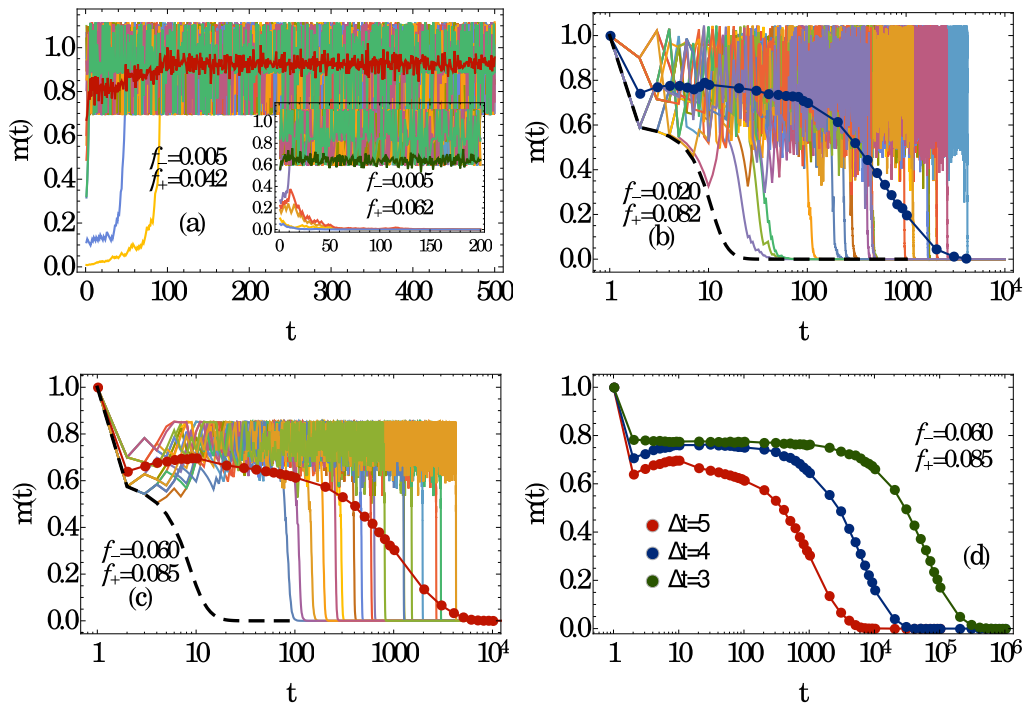}
\caption{MFT temporal disorder analysis: For a RR network with $k_0=12$, $\theta=0.454$ and $\Delta t=5$ the
time evolution of $m$ for distinct sets
of $f_+$ and $f_-$ and distinct independent realizations. Panel $(a)-(c)$ exemplifies
the following cases:  $(f_,f_+)\in(\rm ORD,ME)$ with $\bar{f}<f_b$ and $\bar{f}>f_b$ (inset), 
$(f_-,f_+)\in(\rm ORD,DIS)$ and $(f_-,f_+)\in(\rm ME,DIS)$, respectively. Dashed and symbol curves correspond to the pure versions (for $f=f_+$)and  $m$ averaged over $N_D=10^3$ realizations,
respectively. Panel $(d)$
shows $m$ averaged 
for $\Delta t=3,4$ and $5$.}\label{fig2}
\end{figure}
\end{center}
\end{widetext}

Next, we analyze the cases in which $f_-$ and $f_+$ belong to different phases. Starting with
$f_-\in$ ORD and $f_+\in$ ME (with $f_-<f_b$ and $f_b \leq f_+ \leq f_f$), the phase predominance
can be understood
under a heuristic
analysis, based  on the time evolution for $m(t)<<1$. By recalling that $m(t)$ will increase/decrease exponentially as $m\sim e^{\alpha(f_b-f_-)t}$ and $m\sim e^{-\alpha(f_{+}-f_b)t}$ [with $\alpha$ and $f_b$ given approximately  by Eq. (\ref{largek}) for large $k_0$], respectively, 
the dynamics will be then characterized
for sequences of exponentially increasing and vanishing behaviors, in which
the ordered phase prevails if
$f_++f_-<2f_b$, whereas the metastable phase dominates
when $f_++f_->2f_b$. The line 
fulfilling $f_++f_-=2f_b$ denotes the  crossover between
 ordered and metastable phases lines.

We next consider the case in which
$f_-$ and $f_+$ belong to  the ME and DIS phases,
respectively. Although logistic equations are  different from absorbing phase transitions \cite{fioreneto},
the hysteretic branch makes
the disordered phase  prevailing over metastable one.
Since  the magnetization  vanishes 
irrespectively the initial condition for $f>f_f$, it suffices
  a single long sequence of consecutive (e.g. a rare fluctuation) $f_+$'s in which $m(t)<m_u(f_-)$ for
  the system reaching the disordered phase. For a sufficient long
  time, a rare fluctuation  
  occurs with probability one and thus the temporal disorder will supress the ME phase whenever $f_+<f_f$. A discontinuous phase transition between DIS and ME phases yields at $f_+=f_{f}^-$. 
  Such features are appraised in Figs. \ref{fig2} $(c)$ for
  distinct realizations.
  Despite the prevalence of
  the disorder phase, the average behavior $m(t)$ (measured over many runs) is very different
  from individual runs and it is characterized by a long period
  of a system exhibits ordering until 
  vanishing, as depicted in Figs. \ref{fig2} $(d)$. Note that the time required for the appearance
  of a rare fluctuation increases by lowering $\Delta t$. Thus, such (rare) temporal fluctuations dramatically change the behavior of metastable phase, whose vanishing behavior
  towards the disordered phase is expected to be similar that for APTs  \cite{fioreneto}.

 When $f_-$ and $f_+$ belong to the ORD and DIS phases,
  the resulting phase can also be understood  from the
  competition between deterministic  increasing and vanishing  behaviors 
  at $m(t)<<1$, in similarity with competition between the ORD and ME cases. Thereby, the ordered
and disordered phase will prevail if
${\overline f}<f_b$ and ${\overline f}>f_b$, respectively, where
$2 f_b=f_++f_-$ marks the separatrix between above regimes.
As previously, the average $m(t)$ is  significantly different 
from individual runs and its vanishment also yields for longer times as
$\Delta t$ decreases. These features are exemplified in  
Fig. \ref{fig2}$(b)$ for $f_-=0.020$ and $f_+=0.082$, $N_D=20$ individual runs and $\Delta t = 5$. 
As  it can seen, all realizations and
its average value (symbol curves) remarkably differs from the pure
version (dashed lines).
The prevalence of ORD phase 
is possible only for smaller values of inertia ($1/3<\theta<2/5$ and $3/13<\theta<1/3$ for $k_0=12$ and $k_0=20$, respectively). In particular for $k_0=12$
and $\theta=0.45$, the phase DIS always dominates over the ORD phase, since the lowest $f_-=0$ and $f_+=f_f$ are always greater than $2f_b$. 
\begin{figure}
\includegraphics[scale=0.7]{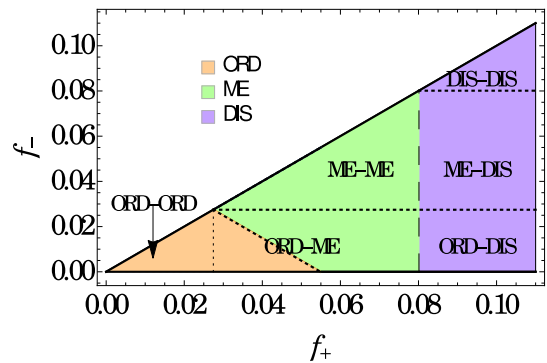}
\caption{MFT phase diagram for RR network with values $k_0=12$ and $\theta=0.45$ under temporal disorder over the control parameter  $f$. The resulting phase is represented by distinct colors. Dotted and dashed lines represent crossovers and discontiuous transition lines, respectively. }\label{fig3}
\end{figure}

From the previous analysis,  we build
the diagram for the temporal disorder IMV for $k_0=12$ and $\theta=0.45$,
as depicted in Fig.\ref{fig3}.   
Dotted and dashed 
lines denote the crossover and phase coexistence lines between ORD/ME and DIS/ME phases, respectively.
Thereby, our findings show that  APT and up-down systems shares  distinct symmetry features, the effect of temporal disorder are similar and are directly related to the bistability of the active/ordered phase.  
                      
\subsection{Temporal disorder in the inertia}
Now we consider the effects of temporal disorder in the inertia, in which its values
are chosen from two possible values $\theta_-$ and $\theta_+$ (with $\theta_+>\theta_-$):
\begin{equation}\label{eq7}
P_{dis}(\theta)=p\delta(\theta-\theta_-)+(1-p)\delta(\theta-\theta_+).
\end{equation}


Although the dependence between the $m(t)$  and
$\theta$ is more cumbersome than the control
parameter $f$, we also consider  Eq. (\ref{eq2}) in the limit of $m(0)<<1$, in which one has a linear equation $dm/dt=A'(f,\theta,k_0)m$.
As mentioned previously, the coefficient
$A'(f,\theta,k_0)$ approaches to Eq. (\ref{largek}) for large $k_0$. 
In particular, $A'(f,\theta,k_0)>(<0)$ according to whether $\theta$ belongs to the ORD (ME/DIS) phases [see e.g. Tables I, II and Eq. (\ref{largek})]. 
For $\theta=\theta_-$ and $\theta_+$,
 $m(t)$ then behaves as $m(t) \sim e^{A'(f,\theta_-,k_0)t}$ and $m(t) \sim e^{A'(f,\theta_+,k_0)t}$, respectively, and thereby the of
 inertial disorder can be analyzed in similarity with the temporal disorder in $f$. The resulting phase then can be predicted from the competition
between distinct behaviors.

Table \ref{t11} and Fig. \ref{fig4}$(d)$ exemplifies coefficients $A'(f,\theta,k_0)$
and the phase diagram
for $f=0.12$ and distinct $\theta$'s, respectively. 
For the pure version, the crossover between ORD and ME phases yields at $\theta_b=1/3$
and ME-DIS discontinous phase transition yields at $\theta_f=2/5$ (see e.g. \ref{fig1}$(a)$).
\begin{table}[h]
\centering
\caption{Coefficients $A'(f,\theta,k_0)$ for $f=0.12$ and $k_0=12$ and the resulting phase.}
\begin{tabular}{c|c|c}
 $\theta$ & $A'(f,\theta,k_0)$&phase  \\ \hline   
0 $<\theta<$1/7 & 0.614...&   ORD     \\
 1/7$<\theta<$ 1/4 &0.467...&ORD   \\
1/4$<\theta<$ 1/3 & 0.192...& ORD\\
1/3 $<\theta<$2/5 &-0.0122...& ME\\
2/5$<\theta<$ 5/11 & -0.0979...&DIS\\
5/11 $<\theta<$1/2& -0.118...&DIS
\end{tabular}
\label{t11}
\end{table}

Starting with $\theta_-$ and $\theta_+$ belonging to the
same phase (ORD/ME/DIS) the resulting phase will be preserved for the temporal disorder, as expected. 
When $\theta_-$ and $\theta_+$ belong to distinct phases, the result phase will depend on the signal of coefficients. 

The case in which $\theta_-$ and $\theta_+$ belong
to ORD and ME/DIS phases, the resulting phase will be
ordered 
if $A'(f,\theta_-,k_0)>A'(f,\theta_+,k_0)$ and ME/DIS
if $A'(f,\theta_-,k_0)<A'(f,\theta_+,k_0)$, respectively. 

The competition  between $\theta_-$ and $\theta_+$ belonging
to the ME and DIS phases will also result in the disordered
phase. Since both $A'(f,\theta_-,k_0)$ and $A'(f,\theta_+,k_0)$
 are negative, the system solely requires 
a long sequence of $\theta=\theta_+$ for driving it to $m(t)<m_u(\theta_-)$ and
then it will evolve to the DIS phase, irrespective the subsequent values of $\theta$. Although more pronounced for $k_0=20$ than for $k_0=12$, but (apparently)
less pronounced than the disorder in the control parameter, 
such case is also featured by a long/remarkable period
  in which the system exhibits ordering until its vanishing
  [see e.g. Figs. \ref{fig4}$(c)$ and \ref{fig45}].
  As previously, a consecutive sequence of $\theta_+$'s
  driving the system to the disordered phase also requires longer
  times for lower $\Delta t$'s and for this reason the time vanishing increases.
A discontinuous phase transition between ME and DIS yields at $\theta_+=\theta_{f}^-$. Thus, the temporal disorder in inertia  also does not suppress
the existence of a discontinuous transition and hysteretic branch.

Since the difference between
the lowest $A'(f,\theta_-,k_0)$ and the largest $A'(f,\theta_+,k_0)$ is always positive,
the phase ORD always prevails over the DIS/ME ones for $k_0=12,f=0.12$ and $p=1/2$, 
(see e.g.   panels $(a)-(b)$ in Fig. \ref{fig4}). The prevalence of the ordered phase
over the disordered and metastable phases in such case
is a new feature originated from the temporal disorder in the inertia, whose
main features
are exemplified in the phase diagram Fig. \ref{fig4}$(d)$.
\begin{table}[h]
\centering
\caption{Coefficients $A'(f,\theta,k_0)$ for $f=0.12$ and $k_0=20$
and the resulting phase.}
\begin{tabular}{c|c|c}
 $\theta$ & $A'(f,\theta,k_0)$&phase  \\ \hline   
3/13 $<\theta<$2/7 & 0.2295...&   ORD     \\
2/7$\le\theta<1/3$ &0.0328.&ORD   \\
1/3$\le\theta<$ 3/8 & -0.0683...& ME\\
3/8  &-0.0876...& ME\\
3/8$<\theta<7/17$&-0.1176..&DIS\\
\end{tabular}
\label{t2}
\end{table}

We close this section by mentioning that  although not presented for $k_0=12$, the competition between ORD and ME/DIS
phases can result to a metastable/disordered as exemplified for $k_0=20$ (see, e.g. coefficients in Table \ref{t2}).

\begin{widetext}
\begin{center}
\begin{figure}
\includegraphics[scale=0.6]{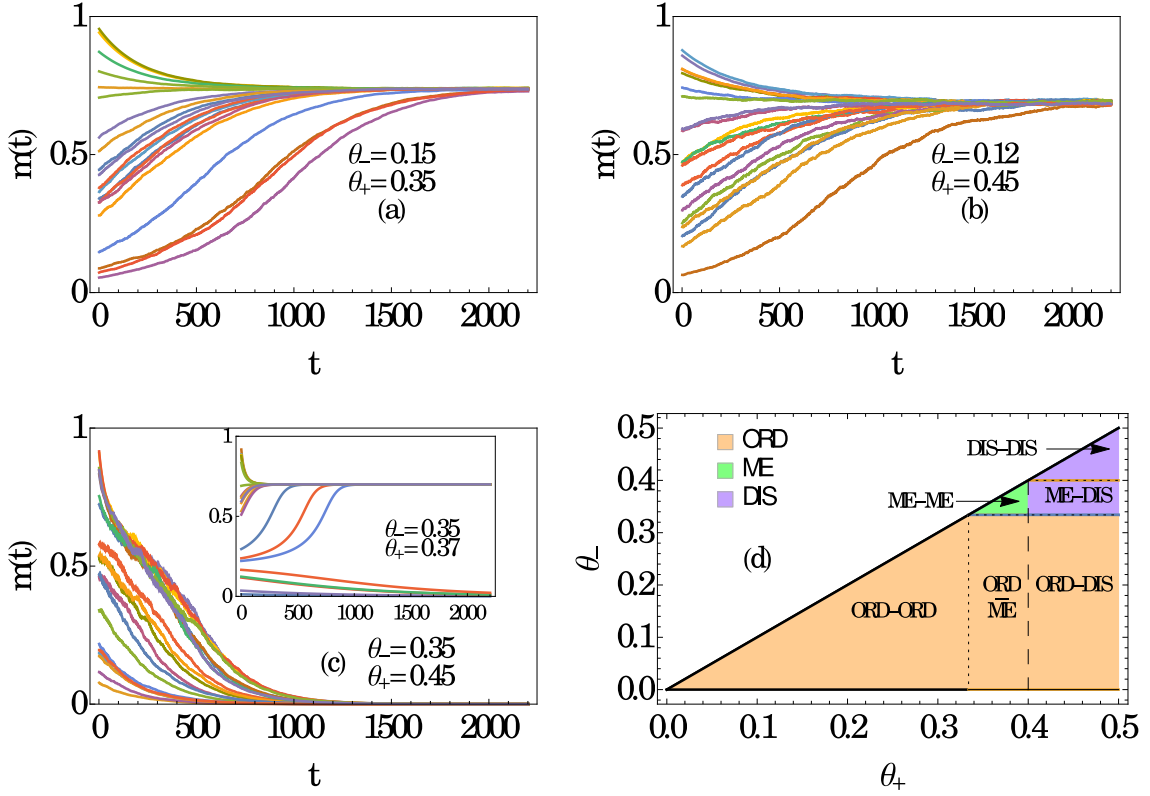}
\caption{MFT analysis for the temporal disorder in the inertia: For RR network with values $k_0=12$ and $f=0.12$,  panels $(a)-(c)$ exemplify the average time evolution of the $m$ for
distinct initial configurations and
sorts of inertia $(\theta_-,\theta_+)\in$: (ORD,ME), (ORD,DIS), (ME,DIS), (ME,ME) [inset], respectively. 
In $(d)$ the  phase diagram with dashed and dotted lines representing discontinuous phase transitions lines and crossover between phases, respectively.
The resulting phase is represented by distinct colors. }\label{fig4}
\end{figure}
\end{center}
\end{widetext}

\begin{figure}
    \centering
    \includegraphics[scale=0.7]{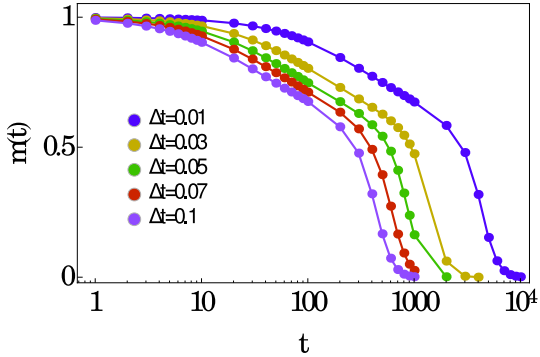}
    \caption{For $k_0=20,f=0.12$, $\theta_-=0.334\in$ ME and 
    $\theta_+=0.412\in$ DIS, the average  $m$
    versus $t$ obtained for $N_D=10^3$
    disorder realizations and distinct 
    $\Delta t$'s.}
    \label{fig45}
\end{figure}

\section{Beyond the mean-field theory: Monte Carlo simulations for distinct kinds of temporal disorder}
In this section, we tackle the influence of temporal disorder beyond the MFT, by analyzing its effect in complex networks structures. We also consider random-regular structures which have been built for fixed connectivity $k_0$ (for a given system size $N$) according to the scheme by Bollob\'as \cite{bollo}. Also,
the neighborhood of each site  has not been altered as the time is changed.

As in the MFT, numerical simulations starts for given initial condition  in which
a new value of the control parameter (whether  $f$ or $\theta$) is   sorted from the two possible values ($f_-/\theta_-$ and $f_+/\theta_+$) for every interval time ranged between $t$ and $t+\Delta t$. The
time evolution of system is analyzed until a maximum time $t_{\rm max}$ that results in a given sequence of  
 $\{f_1,f_2,...,f_M\}$ ($\{\theta_1,\theta_2,...,\theta_M\}$)
 in which $t_{\rm max}=M\times \Delta t$. Such analysis is repeated
 over $N_D=10^3-10^4$ distinct sequences of temporal disorder.
 We have considered $\Delta t=20$ and $t_{\rm max}=10^5-10^6$.

Resulting phases as well as phase transitions can be identified from two distinct (but equivalent) ways.
In the former approach, one considers analysis in the steady state regime in which we start from the ordered phase ($|m|$ close to 1) and  $f$ is raised by an amount $\Delta f$ and the end configuration at $f$ is adopted as the initial condition at  $f + \Delta f$. This procedure is repeat until
the system reaches the disordered phase  at $f_f$. 
Conversely, the numerical simulation is restarted for a given value of
$f$ constrained in the disordered phase but now 
$f$ is decreased by $\Delta f$ until the ordered phase will be reached at $f_b$. Both “forward” and “backward” curves are  expected to coincide
themselves at both ordered and disordered phases, but not along the metastable branch. 

Additionally, the presence of temporal disorder can be more conveniently analyzed (as previously) by 
inspecting the time evolution of order parameter for
distinct initial conditions $0<|m(0)|\le 1$. The system  will converge for a
well defined value  in
both disordered and ordered phases, respectively,
irrespective the initial conditions, whereas it will evolve to two well
defined values  for $f$ constrained in the metastable branch.
Due to the finite size effects, the magnetization never vanishes, but instead, it behaves as $m(t \rightarrow \infty) \sim 1/\sqrt{N}$ in the disordered and metastable phases (for lower $m(0)$).

Although the temporal
disorder features are not expected to
depend on the values of $\theta$ and $k_0$,
the bistable branch is more
pronounced for large connectivities
and  $\theta$'s and
for this reason numerical simulations will be undertaken for $\theta=0.3$ and $k_0=20$,
whose hysteretic loop
for the pure system
was investigated in Ref. \cite{chen2} and 
reproduced in Fig. \ref{fig6}$(a)$. 
As it can be seen, 
for $f<f_b=0.060(5)$ the system is constrained in the ordered
phase, whereas the bistability yields
for $f_b<f<f_f=0.150(5)$. The disordered 
phase emerges for $f>f_f$, irrespective the
initial condition. 
 Fig. \ref{fig6}$(b)-(d)$ depicts the main features
for temporal disorder in the control parameter
for distinct sets of $f_+$ and $f_-$ belonging
to the ORD,ME and DIS phases. In particular, the MFT analysis describes  well the  findings
beyond the MFT, such as the prevalence of the disordered phase over the metastable [Fig.\ref{fig7}$(a)$] for $f_+<f_f^-$ and the competition between ordered and metastable/disordered phases. More specifically, taking into account that
the lowest $f_-$ and $f_+$ are lower than $2f_b$,
the disordered phase always prevails over the ORD one, as illustrated in panel \ref{eq6}(d).
\begin{widetext}
\begin{center}
\begin{figure}
\includegraphics[scale=0.6]{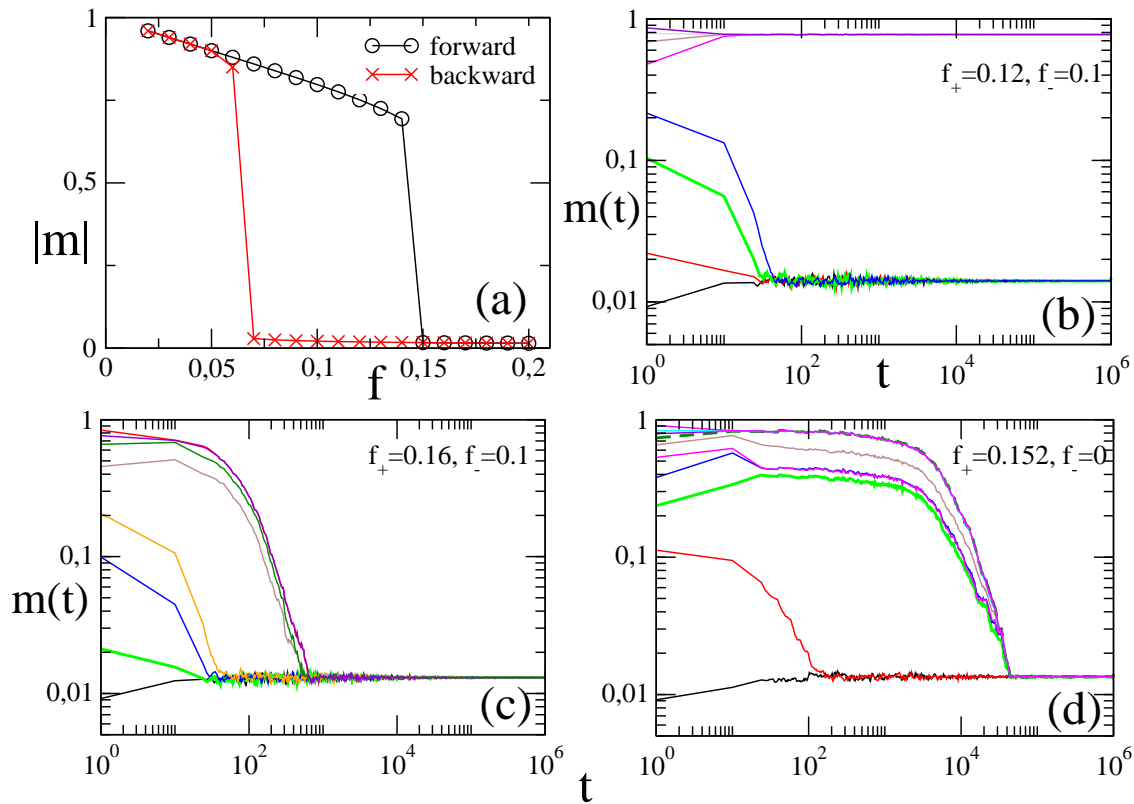}
\caption{Panel $(a)$ depicts, for  $N=5000,k_0=20$ and $\theta=0.3$, the order parameter $|m|$ versus $f$ for the pure system. Continuous and dotted lines denote the forward and backward increase of $f$, respectively. Panels $(b)-(d)$ show the average time evolution of the order parameter $m$ (over $N=10^4$ realizations) for distinct initial conditions $m(0)$ and different sorts of $[f_-,f_+]\in$ [ME,ME],[ME,DIS],[ORD,DIS]. respectively. }\label{fig6}
\end{figure}
\end{center}
\end{widetext}                        
However, due to a finite-size effect the
ORD phase always prevails over the metastable for finite $N$. Since $m(t)$ is finite and proportional to $1/\sqrt{N}$ in 
the disordered phase,  it suffices a long sequence (e.g. a rare fluctuation)
of $f=f_-$ for driving the system to the ORD
phase.  However, such a finite size effect disappears 
as $N\rightarrow \infty$ and MFT also describes well the
prevalence of the ME phase 
when $f_++f_->2f_b$. 
Since the main features are quite similar to those from MFT, we shall omit the phase diagram.
As a final comment, we expect similar trendts for other lattice topologies, although the line separating ordered and other phases does not necessarily will  obey a derivation like MFT.

In the last analysis, we exemplify the main features of
inertia temporal for $f=0.12$. Fig. \ref{fig7}$(b)$ shows the competition between
metastable and disordered phases for $f=0.12$. For
the pure version, the hysteretic branch
is verified for $2/7<\theta\le\theta_f=1/3$ in which the order-parameter jumps
at $\theta>\theta_f$ (see e.g. \cite{jesus2}).
\begin{widetext}
\begin{center}
\begin{figure}
  \epsfig{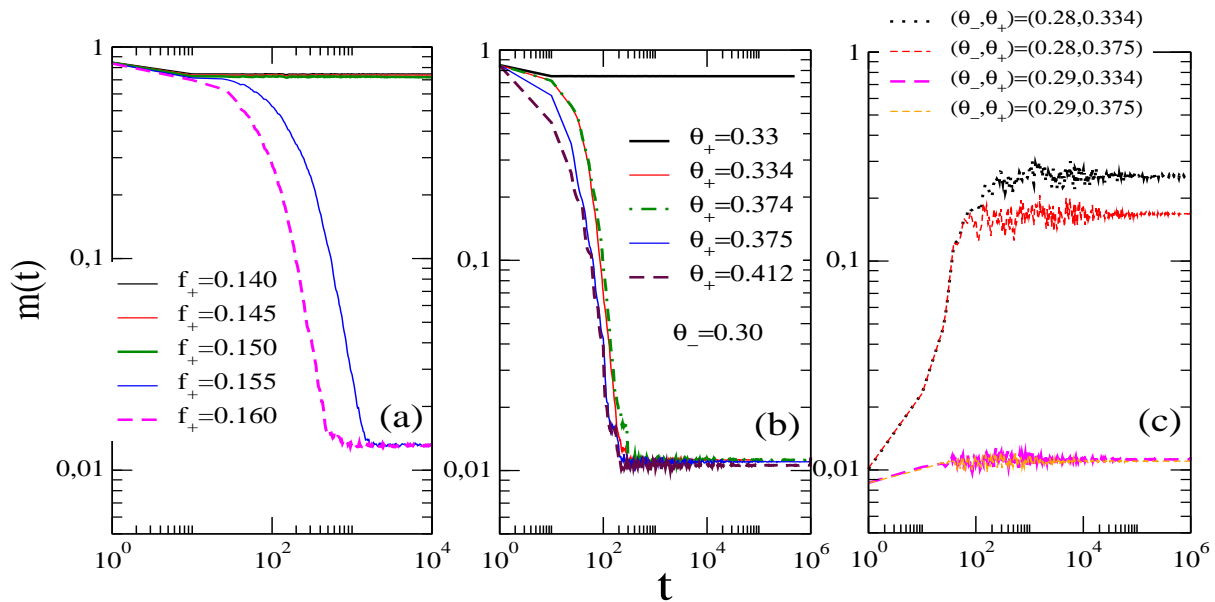}
\caption{For a RR nerwork of size $N=10^4,k_0=20$ and $\theta=0.3$, the average time evolution of the order parameter $m$ 
starting
from the ordered state for $f_-=0.10$ (ME)
for distinct $f_+$'s. Panel $(b)$ depicts, for $f=0.12$ and $m(0)=1$, the time evolution of the average $m$ for $\theta_-=0.30$ (ME)
for distinct $\theta_+$'s. In $(c)$ the same
but for $\theta_-=0.28$ and $0.29$ (both belonging to the ORD phase) and distinct $\theta_+$'s starting from $m(0)=10^{-4}$. In all cases, averages are evaluated over $N=10^3-10^4$ realizations}\label{fig7}
\end{figure}
\end{center}
\end{widetext}
Also in accordance with previous MFT analysis, the competition
between ME and DIS phases always suppresses the phase
coexistence (see e.g. curves for $\theta_+>\theta_f$ in Fig. \ref{fig7}$(b)$)
and a discontinuous transition yields at $\theta_+=\theta_f^-$. On the other hand,
the resulting phase from 
the competition between ORD and DIS phases will depend on particular values of $\theta_-$
and $\theta_+$. More specifically for $\theta _- =0.28$ and $1/3<\theta_+=3/8$
the ORD prevails, whereas the system evolves to disordered phase when 
$\theta _- =0.29$. Since transition points from
 MFT and complex topologies are similar for large  and $k_0$'s
 \cite{chen2}, these above findings can also
 be understood from coefficients
from Table \ref{t2} from which the predominance of  ORD and DIS phase holds for $\theta _- =0.28$
and $0.29$, respectively.

\section{Conclusions}
Based on the MFT, a general description for
discontinuous phase transitions in the presence
temporal disorder was considered. Our theoretical predictions are general
and valid any system displaying a bistable behavior
characterized by the existence of a hysteretic branch. The present study not only confirms
previous findings \cite{fioreneto} but also extends for other  system symmetries and distinct kinds of temporal
disorder. Analysis was exemplified in one of the simplest   "up-down" system symmetry for two kinds of temporal disorder:  inertial majority vote model. 
Since the inertia plays a fundamental role for the emergence of
a discontinuous transition, the effect of its time
variation was also investigated.
Our main findings can be summarized as follows:
Although both kinds of temporal disorder does not suppress existence
of a discontinuous phase transition, the phase coexistence is always supressed when there is
a competition between disordered and metastable
phases. As for with absorbing phase transitions, the competition between different phases can also lead to an order-parameter vanishing characterized by  exponentially large decay times.   The mean-field approach describes very well the effect of temporal disorder in complex topologies.

Our findings are general and expected to be 
valid for other complex structures, such as Erd\"os Renyi and heterogeneous structures.
As a final comment, it will be remarkable to extend such analysis
for discontinuous phase transitions in regular structures, which presents
an entirely different behavior from complex topologies. In theses
systems, ho hysteretic behavior is presented \cite{jesus,fsize2}.

\section{ACKNOWLEDGMENT}
We acknowledge Jos\'e A. Hoyos  for useful suggestions. C. E. F. acknowledges the financial support
from  FAPESP under grant 2018/02405-1 and J. M. E. acknowledges the financial support from CAPES.

\end{document}